\documentclass[11pt]{article}
\newcommand{\s}[1]{{\huge\textsf{\textbf{#1}}}}

\usepackage{amsmath}
\usepackage{amsfonts}
\usepackage[pdftex]{graphicx}
\usepackage{breqn}
\usepackage{float}
\usepackage{pdfpages}
\usepackage{multirow}
\usepackage{longtable}
\usepackage{hyperref}
\hypersetup{colorlinks=true,urlcolor=black, linkcolor=black,citecolor=blue}

\usepackage[utf8]{inputenc}
\usepackage[T1]{fontenc}

\usepackage{mathtools}

\usepackage{epstopdf}
\usepackage{manfnt}
\reversemarginpar
\usepackage{xcolor}
\usepackage{bm}
\usepackage{caption}
\usepackage{subcaption}
\usepackage{wrapfig} 
\usepackage{textcomp}

\def\v   #1{\mathbf{#1}}

\def\sunuc{heterogeneous} 
\def\sunucp{heterogeneous primary} 
\def\prnuc{homogeneous}
\def\prnucp{homogeneous primary}

\graphicspath{{./}}

\title{\s{The role of clearance mechanisms in the kinetics of 
toxic protein aggregates involved in neurodegenerative diseases.} }
\author{
Travis B. Thompson{$^*$}, Georg Meisl{$^\dagger$}, Tuomas Knowles{$^{\dagger,\ddagger}$},
 and Alain Goriely{$^*$}\\ \\
$^*$\small{Mathematical Institute, Andrew Wiles Building}\\
\small{Woodtsock Rd University of 
Oxford OX2 6GG, UK}\\
$^\dagger$\small{Centre for Misfolding Diseases, Department of Chemistry,}\\
\small{ University of Cambridge, Lensfield Road, Cambridge CB2 1EW, UK}\\
$^\ddagger$\small{Cavendish Laboratory, University of Cambridge,}\\
 \small{19 JJ Thomson Avenue, Cambridge CB3 0HE, UK }
}

\begin{document}
\maketitle
\begin{abstract}\vskip 6pt
\hrule\vskip 12pt
Protein aggregates in the brain play a central role in  
cognitive decline and structural damage associated with neurodegenerative 
diseases.  For instance, in Alzheimer's disease the formation of Amyloid-beta plaques and  tau proteins neurofibrillary tangles  follows from the accumulation of different proteins into large aggregates through {specific mechanisms} such as nucleation and elongation.  These 
mechanisms have been {studied} in vitro 
where total protein mass is  conserved. However, in vivo, clearance mechanisms may play an important role in limiting the formation of aggregates. 
Here, we generalize classical models of protein aggregation 
to take into account both  {production} of monomers and the clearance of 
protein aggregates.  Depending on the clearance model, we show that  there may be a critical
 clearance value above which aggregation does not take place.  Our result offers 
further evidence in support of the hypotheses that clearance mechanisms play 
a potentially crucial role in neurodegenerative disease initiation and 
progression; and as such, are a possible  therapeutic 
target. \vskip 12pt
\hrule
\end{abstract}

\section{Introduction}
Alzheimer's disease (AD), and other related neurodegenerative diseases, are associated with 
the assembly of specific, toxic proteins into fibrillar aggregates.  
Alzheimer's disease, in particular, is characterized by the aggregation of Amyloid-$\beta$ 
(A${\beta}$) plaques and tau protein neurofibrillary tangles (NFT).  The role of A${\beta}$ in Alzheimer's is thought to be so central to the 
disease that it is the basis of the so-called `Amyloid-$\beta$ hypothesis' \cite{hardy1992alzheimer,hardy1991amyloid,selkoe2016amyloid}, stating that 
that the accumulation and deposition of oligomeric or fibrillar 
amyloid beta peptide is the main cause of the disease.  This hypothesis has 
provided a guide for most of AD research over the last 20 years.  
However, recent experimental evidence, and the failure of several 
drug trials, has lead to renewed scrutiny  of this foundational assumption. 

The production of A${\beta}$ is a natural process related to neuronal 
activity. Indeed, A${\beta}$ is a normal metabolic waste byproduct \cite{bacyinski2017,benveniste2019} that is
typically removed from intracellular and extracellular compartments by several
clearance mechanisms \cite{tarasoff2015,Xin2018}. In  healthy subjects
waste proteins are broken down by enzymes, removed by cellular uptake, or 
efflux to cerebrospinal fluid compartments where they  eventually reach 
arachnoid granulations, or lymphatic vessels.  While healthy clearance mechanisms, 
working in harmony, avert the buildup of toxic A${\beta}$ 
plaques and tau NFT; their impairment or dysfunction can lead  to toxic levels 
of aggregates \cite{Xin2018}.  %
The specifics of in-vivo clearance mechanisms 
remain a  topic of clinical debate; however, the kinetics 
enabling proteins to amass into toxic aggregates can be carefully, and 
systematically, studied in vitro and under varied conditions.  The production 
of A${\beta}$, at a high level, is mediated by a membrane protein called 
amyloid precursor protein (APP).  APP is typically cleaved by $\alpha$-secretase and the resulting products do not aggregate. However, {APP can also be cleaved by 
$\beta$-secretase}, which results in soluble 
monomeric APP fragments of different sizes. {The most common size categories are 
} A$\beta$38, A$\beta$40, and A$\beta$42.  While monomeric A$\beta$38 is 
not prone to further aggregation;  A$\beta$40 and A$\beta$42, 
{containing} two additional amino acids at the C terminus, are the main 
{isoforms} of {interest in the study of AD pathology}.

Protein aggregation pathways are, in general, complex and involve multiple 
steps.  In fact, it has recently been shown \cite{meisl2014differences} that 
the aggregation properties of A$\beta$40, which is more abundant, differ 
from those of the more aggregate-prone A$\beta$42; even under the same 
conditions.  A theoretical framework of chemical kinetics and aggregation 
theory \cite{cohen2011nucleated,cohen2011nucleated2,cohen2011nucleated3} has 
been combined with careful, systematic in vitro experiments performed 
under differing conditions; such as varied concentration or pH.  {This 
approach has: elucidated effective pathways} and mechanisms for nucleation, 
aggregation and fragmentation \cite{meisl2016molecular}; and produced a deep 
understanding of key properties, underlying the formation of aggregates under 
ideal conditions, with the potential for {therapeutic} intervention 
\cite{frankel2019autocatalytic,kundel2018measurement}.

Here, we develop a mathematical framework to describe the effects of clearance and 
monomer production chemical kinetics driving aggregation; we apply the framework to 
the study of A${\beta}$.  To accomplish this, we extend the current theory describing  
A${\beta}$ aggregation in vitro, which has been validated against experiment, to 
include monomer production and oligomer clearance terms.  In particular, we study two different clearance 
mechanisms: one where total mass is conserved (size-independent clearance); and 
one where it is not (size-dependent clearance). In the former case we show the 
full system reduces to three equations amenable to a systematic analysis.  We 
 identify a critical value of clearance above which the production of toxic 
aggregates does not take place.  Our results offer further evidence in support of two main 
hypotheses: that clearance mechanisms play a crucial role in 
{neurodegenerative} disease initiation and progression; and that 
therapies enhancing clearance above a prescribed, critical value may serve as 
a possible intervention strategy.  In particular, we will exhibit the existence 
of critical clearance values; such values are consistent with the observation of 
disease onset when natural clearance mechanisms within the brain have degraded 
through aging. 

\section{{A model of toxic protein aggregation}}
\label{sec:model}
\begin{figure}[ht]
\centering \includegraphics[width=.9\linewidth]{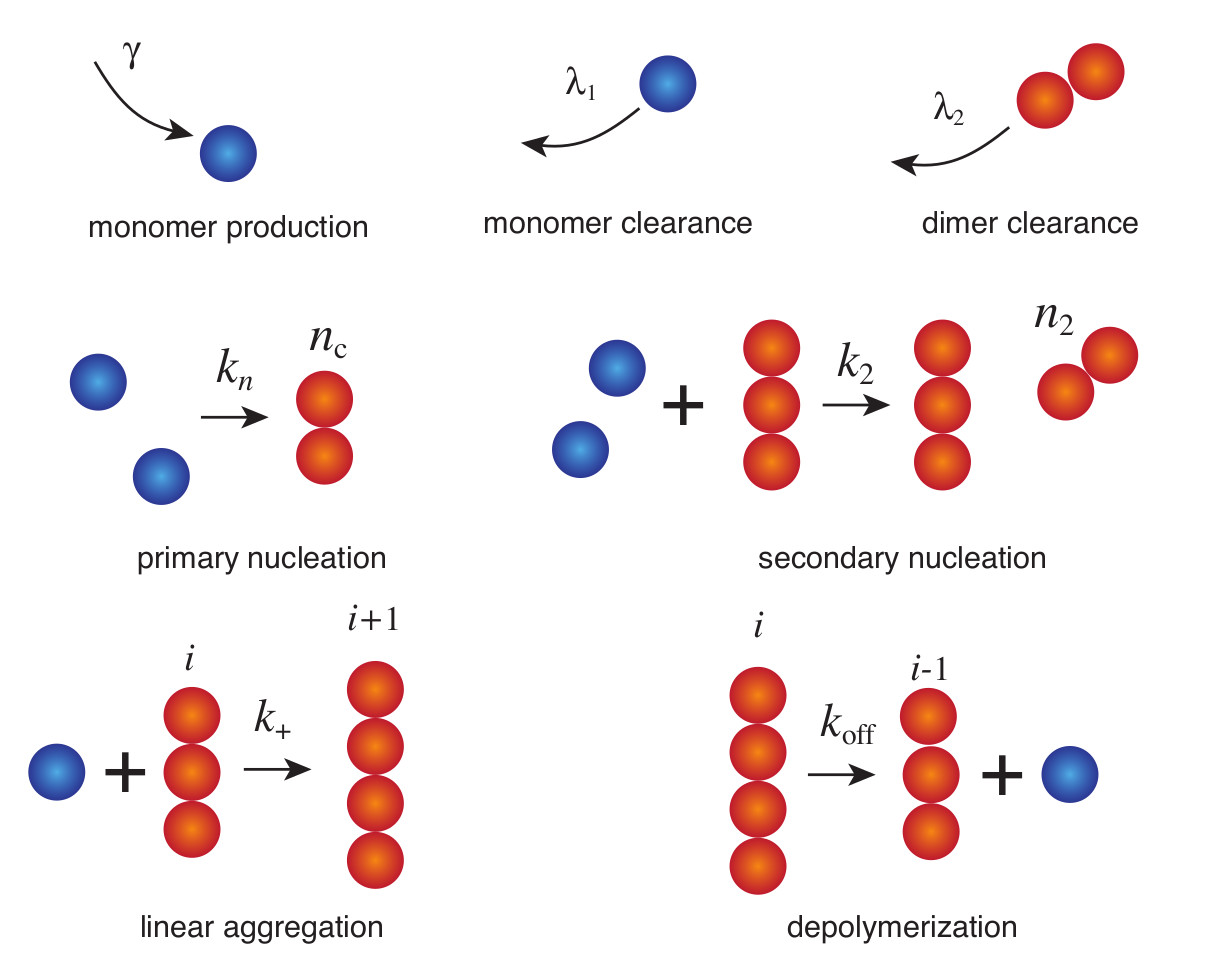}
\caption{Mechanisms included in the master equations \eqref{mbeta1}-\eqref{mbeta5}.
We consider multiple effects for the formation 
of aggregates into our systems with rates constants $k_{i}$. The constants 
corresponding to transfer of mass to and from the external system are represented by greek 
letters ($\gamma$ and $\lambda_{i}$). 
The process of {\sunuc} nucleation (with constant $k_{0}$) is 
similar to {\prnucp} nucleation and is not depicted (the main difference being that its 
rate does not depend on the monomer concentration).}
\label{Fig-aggregation}
\end{figure}

Our model for protein aggregation-dynamics  
model includes multiple mechanisms: 
{\sunucp} nucleation; {\prnucp} nucleation; secondary nucleation; linear elongation; fragmentation; and 
clearance (c.f.~Fig.~\ref{Fig-aggregation}).  These mechanisms lead to a general class of 
mathematical models that can describe a wide range of aggregating systems in vitro.  In particular, 
by including {\sunucp} nucleation terms, a source term for new nuclei, that is independent 
of monomer concentration, is present; this source is in addition to the usual monomer-dependent 
{\prnucp} nucleation.  Thus, in such a model, the importance of interfaces in the initiation of 
nucleation is sufficiently accounted for.  In the model, each aggregate of a given size is represented 
by a population.  In general, each population, with aggregates of size $i$, will be represented by 
an indexed concentration; we use the special notation $m(t)$ for the monomer population 
$i=1$, while all other aggregate concentrations are denoted by $p_i(t)$ for $i\in \left\{2,3,\dots\right\}$.  %
%
The master equations are then:
\begin{eqnarray}\label{mbeta1}
&&  {\frac{{\text{d}}{{m}}}{{\text{d}}t}}  
=\gamma - \lambda_1 m-2 k_{0}-n_{c} k_{n} m^{n_{c}}-2 k_{+}m P-n_{2} %
k_{2} \sigma(m) M+2 k_{\text{off}}\,P
\\ \label{m2}
&&  {\frac{{\text{d}}{{p_2}}}{{\text{d}}t}}  
= - \lambda_2 p_2+k_{0} +\delta_{{2},n_{c}}k_{n} m^{n_{c}}-2 k_{+}m %
p_{2}+\delta_{2,n_{2}} k_{2} \sigma(m) M+2 k_{\text{off}}\, p_{3}\\
&&  {\frac{{\text{d}}{{p_i}}}{{\text{d}}t}}  
=  - \lambda_i p_i+\delta_{i,n_{c}}k_{n} m^{n_{c}} + %
2 k_{+}m (p_{i-1}-p_{i})+2 k_{\text{off}} (p_{i+1}-p_{i}) + %
\delta_{i,n_{2}} k_{2} \sigma(m) M,\quad i>2,
\label{mbeta5}
\end{eqnarray}
where $\delta_{i,j}$ is Kronecker's delta (1 if $i=j$ and 0 otherwise) and
\begin{equation}\label{eqn:moments}
\sigma(m) = \frac{m^{n_2} K_M }{K_M + m^{n_2}}, \quad %
P=\sum_{i=2}^{\infty} p_{i}, \quad  M=\sum_{i=2}^{\infty} i p_{i}.
\end{equation}
Here, $P$ and $M$ are the first two moments 
of the population distribution; they represent the total number and total mass 
of aggregates, respectively. 
In these equations,  the parameters  represent the following effects,  
sketched in Fig.~\ref{Fig-aggregation}:
 $\gamma$: (constant) monomer production such as by 
$\beta$-secretase mitigated cleavage of APP, driving mass influx;
 $\lambda_{i}$: clearance of aggregate of size $i$  such as by lymphatic or 
cellular processes;
 $k_{0}$: {\sunucp} nucleation (independent of the monomer concentration);
 $k_{n}$ and $n_{c}$: nucleation of aggregates of size $n_{c}>1$;
 $k_{+}$: linear elongation transforming  aggregate from size $i$ to $i+1$;
 $k_{2}$: secondary nucleation of aggregates of size $n_{2}>1$;
 $K_{M}$: saturation of the secondary nucleation;
 $k_{\text{off}}$: depolymerization by one monomer.

{The aggregation model \eqref{mbeta1}-\eqref{mbeta5}, and its many variations, have served as a 
template for in vitro experiments
\cite{meisl2014differences,frankel2019autocatalytic,cohen2013proliferation}}.  
Multiple experimental fittings have shown that the exponents $n_{c}$ and $n_{2}$ are: $n_c = n_2 = 2$ for A$\beta$40 and for A$\beta$42 in the presence of a PBS buffer 
\cite{meisl2014differences,cohen2013proliferation}; for A$\beta$42 in the 
presence of a HEPES buffer $n_2 = 2$ and $n_c = 0$ provides the best fit 
\cite{frankel2019autocatalytic}.  {For 
the discussions and derivations in this manuscript we take the view of 
PBS buffer experiments \cite{meisl2014differences,cohen2013proliferation} 
so that $n_c = n_2 = 2$}.  {Adaptation to A$\beta$42 HEPES, so that 
$n_c = 0$, is straightforward and all numerical results are qualitatively 
similar.}  
When fitting experimental data it is often 
the case that only one of $k_0$ or $k_n$, depending on the best data fit, is 
used; i.e.~that either {\sunucp} nucleation or {\prnucp} nucleation 
{best explains the particular experimental data.

The primary purpose of 
this manuscript is to describe the qualitative impact of clearance mechanisms 
in the dynamics and a particular choice of nucleation mechanism, i.e.~$k_0=0$ versus $k_n=0$, does not 
affect the results.}  %
Examples of {fitted} A$\beta$ model parameters {are} listed in 
Table~\ref{tab:parameters}.  PBS and HEPES refer to the buffers used in the 
corresponding experiments.  %
Aggregation, in the fitted experiments, {proceeds}  much faster than  depolymerization and $k_{\text{off}}=0$ is found to be a good fit to describe the dynamics. However, from a theoretical point of view, we note that $k_{\text{off}} = 0$ implies that there is no non-vanishing stationary distribution in the absence of clearance and production terms. Here, we will first follow experimental data and take  $k_{\text{off}} = 0$. Then, we will show that the addition of this small term does not change our results. Therefore, we will use the fitted experimental parameters given in  Table~\ref{tab:parameters}.  Clearance and production have not been investigated experimentally; thus, we leave them as free parameters. In particular, we will be interested in determining particular values of these parameters when a qualitative change of the dynamics occurs.

\begin{table}[h]
\caption[Kinetic parameters]{Typical  parameters for the  $A\beta$ model.  PBS and HEPES refer to the buffer 
used for the experiments. Note that A$\beta$42 is generally faster than 
A$\beta$40 and HEPES buffer is faster than PBS. In these experiments, 
aggregation is sufficiently fast so that $k_{\text{off}}=0$ provides a good fit.  The 
values of $\lambda_{\text{crit}}$  give the critical values of clearance and 
their approximations for the size-independent case (see text). The values of 
$\tau_{1}$ and $\tau_{2}$ give the typical time scales associated with each 
dynamics (see text). }
      \centering
        \begin{tabular}{l|l|c|c|c|l}
\hline
param.& mechanism& A$\beta$40 PBS \cite{cohen2013proliferation} & A$\beta$42 PBS \cite{meisl2014differences}& A$\beta$42 HEPES \cite{linse2019kinetic}&units\\
 \hline\hline
$k_0$ & {\sunuc} nucleation      	& 0				  & 0				& $1.6 \times 10^{-11}$ &M\,h$^{-1}$\\
$k_n$ & {\prnuc} nucleation	& $5.8 \times 10^{-3}$ & $1.2 \times 10^{-1}$	& 0&M$^{1-n_c}$h$^{-1}$\\
$n_c$ & {\prnuc} nucleation 	& 2				& 2				& 2&unitless	\\
$k_2$ & secondary nucleation 	&$1.1 \times 10^{7}$	& $3.6 \times 10^{7}$	& $2.1 \times 10^{14}$&M$^{-2}$h$^{-1}$\\
$n_2$ & secondary nucleation 	&2				& 2				& 2	&unitless					\\
$K_M$ & saturation      	&$3.6 \times 10^{-11}$	& $3.6 \times 10^{-12}$	& $2.3 \times 10^{-17}$&M$^{2}$\\
$k_+$ & elongation  			   & $1.1 \times 10^{9}$	& $1.1 \times 10^{10}$	&  $1 \times 10^{10}$  &M$^{-1}$h$^{-1}$\\
$k_{\text{off}}$ & depolymerization  & 0                                  & 0    	                         &0                                 & h$^{-1}$\\
$m_0$ &Initial monomer c. 	& $3 \times 10^{-6}$& $3 \times 10^{-6}$& $3 \times 10^{-6}$&M  \\
 \hline\hline
 $\lambda_{\text{crit}}$ &critical clearance 	& 0.72& 2.45&17.0 &h$^{-1}$  \\
  $\tilde{\lambda}_{\text{crit}}$ &perfect bifurcation 	& 0.72& 2.47&17.0 &h$^{-1}$  \\
  $\alpha$&nonlinear coefficient &312,042&647,390&2.83726$\times$10$^6$&M$^{-1}$ h$^{-1}$ \\
   \hline\hline
 $\tau_{1}$ & exponential time scale 	&1.4&0.4&0.06 &h  \\
  $\tau_{2}$ & amplification time scale 	& 12.6& 2.5&0.4 &h  \\
   \hline\hline
     $\lambda_{\text{crit}}^{(1)}$ &critical clearance $\nu=1$	& 7.8$\times 10^{-5}$&9.2$\times 10^{-5}$&4.8$\times 10^{-3}$ &h$^{-1}$  \\
 $\lambda_{\text{crit}}^{(0)}$ &critical clearance $\nu=0$	& 0.72& 2.47&17.0 &h$^{-1}$  \\
     $\lambda_{\text{crit}}^{(-1)}$ &critical clearance $\nu=-1$	& 13.2$\times 10^{3}$&13.2$\times 10^{4}$&12$\times 10^{4}$ &h$^{-1}$  \\
\end{tabular}
        \label{tab:parameters}
\end{table}

\section{{Size-independent clearance}}
\label{sec:size-indep-clearance}
In the case of size-independent clearance, we have $\lambda_{i}=\lambda>0$ for all $i$.
Our main question is to understand the role of the clearance term.  In particular, we will establish that if clearance is sufficiently large,  the formation of aggregates does not take place.

\subsection{Moment analysis}
In the size-independent case, a  well-known but remarkable feature of the system \eqref{mbeta1}-\eqref{mbeta5} is that a  closed system of equations for the first two moments $P$ and $M$ and the monomer concentration $m=p_{1}$ can be obtained exactly: 
\begin{eqnarray}
&&{\frac{{\text{d}}{{P}}}{{\text{d}}t}}  
\,\,= \phantom{\gamma}- \lambda P+ k_{0}+ k_{n} m^{2}+ k_{2}\, \sigma(m) M,
 \label{P}\\
&&{\frac{{\text{d}}{{M}}}{{\text{d}}t}}  
=\phantom{\gamma}- \lambda M+2 k_{0}+{2( k_{+}m-k_{\text{off}})P} +2 k_{n} m^{2}+ {2}k_{2}\, \sigma(m) M,
 \label{M}\\
&&  {\frac{{\text{d}}{{m}}}{{\text{d}}t}}  
\,= \gamma- \lambda m\,-\,2 k_{0}-{2( k_{+}m-k_{\text{off}})P}-2 k_{n} m^{2}-{2} k_{2}\, \sigma(m) M,
 \label{m}
\end{eqnarray}
where 
$\sigma(m)={m^{{2}}K_{M}} /({K_{M}+{m^{{2}}}})$ and we have chosen $n_{2}=2$. 
The total mass of the system  $M_{\text{tot}} = M + m$ satisfies, by summing \eqref{M}-\eqref{m}, the evolution equation 
\begin{equation}\label{eqn:Mtot}
{\frac{{\text{d}}{{M_{\text{tot}}}}}{{\text{d}}t}} = -\lambda M_{\text{tot}} + \gamma. 
\end{equation}
This equation implies that the total mass in the system evolves to a stable steady state $M_{\text{tot}}= \gamma/\lambda$ with a typical time-scale $1/\lambda$. To simplify the analysis, we will further assume that, initially, the system is at this state by choosing the following {unseeded initial conditions}
\begin{equation}
M(0)=P(0)=0,\quad m(0)=m_{0}=\gamma/\lambda,
\end{equation}
and the total mass of the system is conserved for all time $M_{\text{tot}}(t)=m_{0}$. The term `unseeded' refers to the fact that, initially, there is no toxic protein in the system (hence, no seed). 
This condition assumes  {a 
lack of aggregated species} in a healthy in vivo state. Indeed, it is observed that 
soluble A$\beta$ monomers are found in healthy individuals of all ages while 
aggregates larger than monomers are correlated with Alzheimer's disease 
progression \cite{brody2017}.  An extra advantage of this approach is that it fixes the constant $\gamma=m_{0}\lambda$. 

\begin{figure}[ht]
\centerline{\includegraphics[width=0.8\linewidth]%
{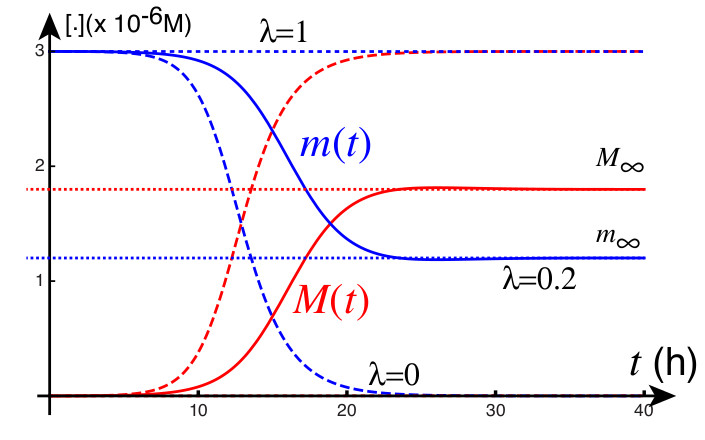}}
\caption{\label{FigmM-sizeindep2} Typical dynamics of the monomer (blue) and toxic (red) concentration (in moles) for different values of the clearance  ($\lambda$ in h$^{-1}$) for A$\beta$40 with parameters from Table 1 and $\lambda=0$ (large dashed), $\lambda=0.2$ (solid) and $\lambda=1$ (small dashed). Asymptotic values for $\lambda = 0.2$ are shown with dotted lines.} 
\end{figure}

 Before we study the system in full generality,  it is useful to consider the overall  dynamic of the system for a typical set of parameters for the aggregation of  A$\beta$40 given in the first column of Table~1. We will use this set of parameters for all our examples. The other data sets are qualitatively equivalent and the values of various derived quantities are given in Table 1. As shown in Fig.~\ref{FigmM-sizeindep2}, the typical behavior of the system from an unseeded initial condition is for the toxic protein mass to increase up to  finite value $M_{\infty}$ while the monomer concentration decreases to  $m_{\infty}$ in a typical sigmoid-like behavior.  
We observe that, in the absence of clearance, the 
monomer population is completely converted to toxic proteins ($\lambda=0$, 
dashed curves in Fig.~\ref{FigmM-sizeindep2}). Conversely, for large clearance 
almost no conversion takes place ($\lambda=1$, dotted curves %
in Fig.~\ref{FigmM-sizeindep2}).   Some of the monomers are converted 
(solid curves for $\lambda=0.2$ in Fig.~\ref{FigmM-sizeindep2}) for the case of 
moderate clearance.   Of particular interest for our discussion is the change of behavior at some  critical value $\lambda_{\text{crit}}$ of the clearance $\lambda$ where aggregation becomes negligible. 

To derive an exact value for $\lambda_{\text{crit}}$, we determine
the dependence of the asymptotic states $m_{\infty}$ on 
$\lambda$.  Using the steady state hypothesis with $m=m_{\infty}$, 
$P = P_{\infty}$ and $M = M_{\infty}$ in \eqref{P}-\eqref{M} one expresses the 
latter two states as a function of the parameters, $\lambda$ and $m_{\infty}$.  
These relations are substituted in \eqref{m} to produce the implicit equation  
$q(\lambda,m_{\infty}) = 0$ with
\begin{dmath}\label{eqn:q-lamcrit}
q(m_\infty,\lambda)=2 k_+ k_n m_{\infty }^5
-m_{\infty }^4 \left(2 k_+ k_2 K_M-2 \lambda  k_n+2 k_n k_{\text{off}}\right)
-m_{\infty }^3 \left(-\lambda
   ^2-2 k_+ k_2 m_0 K_M+2 k_2 \lambda  K_M-2 k_+ k_n K_M-2 k_2 k_{\text{off}} K_M-2 k_+
   k_0\right)-m_{\infty }^2 \left(-2 k_0 \lambda -2 k_2 \lambda  m_0 K_M+2 k_2 m_0 k_{\text{off}}
   K_M-2 \lambda  k_n K_M+2 k_n k_{\text{off}} K_M+2 k_0 k_{\text{off}}+\lambda ^2
   m_0\right)+m_{\infty } \left(\lambda ^2 K_M+2 k_+ k_0 K_M\right)+2 \lambda 
   k_0 K_M-2 k_0 k_{\text{off}} K_M-\lambda ^2 m_0 K_M
    \end{dmath}

\begin{figure}[ht]
\centerline{\includegraphics[width=0.9\linewidth]%
{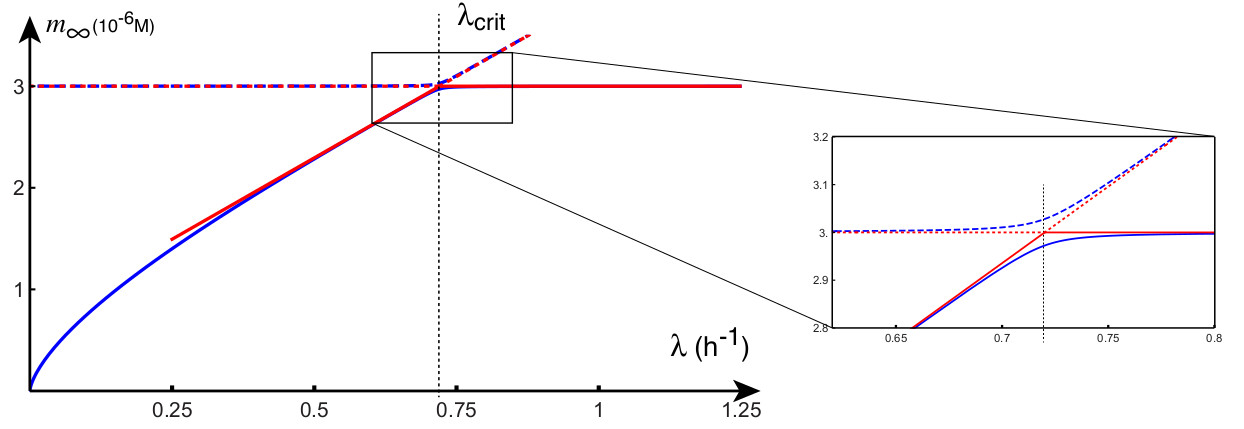}}
\caption{\label{Fig-bif1} %
Perfect (red) and imperfect (blue) transcritical bifurcation obtained for %
A$\beta$40. Unstable (dashed) and stable(solid) equilibrium solutions.
In this case, we have $\tilde{\lambda}_{\text{crit}}\approx 0.72$ and 
$\tilde{m}_{\infty}\approx 0.7 + 3.2 \tilde{\lambda}$. Dashed curves 
indicate unstable equilibria solutions and solid curves denote stable 
equilibria.
}
\end{figure}

For instance, for the same parameter values as in Fig.~\ref{FigmM-sizeindep2}, we show in Fig.~\ref{Fig-bif1} the values of $m_\infty$ as a function of $\lambda$. We observe a sharp transition for a critical value of the clearance parameter $\lambda$. 

There are three necessary conditions for $\lambda_{\text{crit}}$: first 
that $\lambda_{\text{crit}}$ is non-negative; second that $m_{\infty}$ is maximal; 
and third that the value of $m_{\infty}$ coincides with $m_0$.  The last two 
conditions can be realized by computing the derivative of the expression 
$q(\lambda,m_\infty)=0$ evaluated at $m = m_{\infty}$.  Therefore, $  \lambda_{\text{crit}}$ is given by the positive root of  $L(\lambda)=0$ where
\begin{dmath}\label{eqn:l-lamcrit}
L(\lambda) = \left.\frac{\partial q}{\partial m_\infty}\right\vert_{m_\infty=m_0}%
= -4 k_{\text{off}} K_M m_{\infty } \left(k_2 m_0+k_n\right)+6 k_+ k_n K_M
   m_{\infty }^2+6 k_2 k_{\text{off}} K_M m_{\infty }^2-8 k_+ k_2 K_M m_{\infty }^3+6 k_+ k_2 m_0 K_M
   m_{\infty }^2+2 k_+ k_0 K_M-8 k_n k_{\text{off}} m_{\infty }^3+10 k_+ k_n m_{\infty }^4-4 k_0
   k_{\text{off}} m_{\infty }+6 k_+ k_0 m_{\infty }^2+\lambda  \left(4 K_M m_{\infty } \left(k_2 m_0+k_n\right)-6 k_2 K_M m_{\infty }^2+8 k_n m_{\infty }^3+4
   k_0 m_{\infty }\right)+\lambda ^2 \left(K_M+3 m_{\infty }^2-2 m_0
   m_{\infty }\right)
   \end{dmath}
For A$\beta$-40 the critical clearance, as shown in  Fig.~\ref{Fig-bif1},  is $\lambda_{\text{crit}} = 0.72$. Critical clearance rates for the other experimental data sets are given in 
Table~\ref{tab:parameters} for comparison.

\subsection{Bifurcation and normal form analysis}
In a  neighborhood of $\lambda_{\text{crit}}$, $m_{\infty}$, as a function of $\lambda$, undergoes a sharp transition.  This transition is not a 
bifurcation in the strict sense but, in the parlance of dynamical systems, it can be described as an imperfect transcritical bifurcation when  {\sunuc} nucleation and {\prnuc} nucleation terms can be understood as an imperfection and are
sufficiently small with respect to the elongation. 
More specifically, when  $k_{0}/(k_{+} m_{0}^{2})\ll 1$ and ${k_{n}}/k_{+}\ll1$ 
the system is well approximated by $k_{0} = 0$ and $k_n = 0$. In this limiting case, the fixed point $(P,M,m) = (0,0,m_0)$ for the system \eqref{P}-\eqref{m}  undergoes a (perfect) transcritical bifurcation    at $\tilde{\lambda}_{\text{crit}}\approx {\lambda}_{\text{crit}}$ that can be obtained by locally expanding $m_{\infty}$ in $\lambda$ to find
\begin{equation}
\tilde m_{\infty}=m_{0} + %
\frac{1}{\alpha} (\lambda-\tilde{\lambda}_{\text{crit}})+ %
\mathcal{O}\left((\lambda-\tilde{\lambda}_{\text{crit}})^{2}\right),
\end{equation}
where $\tilde{\lambda}_{\text{crit}}$ is specified by the formula
\begin{align}
&\tilde{\lambda}_{\text{crit}}=
\frac{m_0 \left(\sqrt{k_2 K_M \left(m_0 \left(k_2 m_0+2 k_+\right) K_M-2 k_{\text{off}} K_M+2 m_0^2
   \left(k_+ m_0-k_{\text{off}}\right)\right)}+k_2 m_0 K_M\right)}{K_M+m_0^2},
 \end{align}
and ${\alpha}$ is defined by the expression 
\begin{align}\label{alpha}
   &\alpha=\frac{m_0 \left(k_2 K_M \left(2 \tilde{\lambda}_{\text{crit}}+3 k_+ m_0-2 k_{\text{off}}\right)-\tilde{\lambda}_{\text{crit}}^2\right)}{ \tilde{\lambda}_{\text{crit}} \left(K_M+m_0^2\right)-k_2 m_0^2 K_M}.
\end{align}
{When the clearance is close to the critical value} the linear approximation 
to the perfect bifurcation is a reasonable approximation for the imperfect 
bifurcation as can be appreciated in Fig.~\ref{Fig-bif1} where %
$\tilde{\lambda}_{\text{crit}}\approx 0.72$ and %
$\tilde{m}_{\infty}\approx 0.7 + 3.2 \tilde{\lambda}$.  
By analogy with epidemiology we define a dimensionless \textit{neurodegenerative reproduction number}
\begin{equation}
R_{0}=\frac{\tilde{\lambda}_{\text{crit}}}{\lambda},
\end{equation}
such that for $R_{0}<1$ the protein toxic level is negligible and grows to finite value for $R_{0}>1$.

The existence of a critical clearance rate shows that in the healthy regime, %
i.e.~for sufficiently large values of clearance, the system 
\eqref{mbeta1}-\eqref{mbeta5} with size-independent clearance can support a 
small, endemic, population of toxic proteins.  {The aggregation of a 
significant toxic population, in this case, occurs only when the system's 
clearance rate, $\lambda$, drops sufficiently below the critical clearance 
rate $\lambda_{\text{crit}}$}.  %
We can explore the dynamics close to the bifurcation by considering the normal form of the system for the perfect system (\eqref{P}-\eqref{m} with $k_{0} = k_n = 0$) near $\lambda = \tilde{\lambda}_{\text{crit}}$. The general method to obtain the normal form of a transcritical bifurcation for an arbitrary smooth vector field is given in  Appendix~\ref{appdx:a}. Applying these ideas, we can approximate the full system by 
\begin{align}
&\dot P=-(\lambda-\tilde{\lambda}_{\text{crit}}) P+\frac{\alpha}{v_P} P^2,\label{eqn:nrmfrm:P}\\
&\dot M=-(\lambda-\tilde{\lambda}_{\text{crit}}) M-{\alpha} M^2,\phantom{\frac{\alpha}{v_P}}\label{eqn:nrmfrm:M} \\
&\dot m=-(\lambda-\lambda_{\text{crit}}) (m-m_0)+{\alpha} (m-m_0)^2,\label{eqn:nrmfrm:m}
\end{align}
where $\alpha$ is given by \eqref{alpha} and
\begin{equation}
v_P=-\frac{ \tilde{\lambda}_{\text{crit}} }{2 \left(k_+ m_0-k_{\text{off}}+ \tilde{\lambda}_{\text{crit}}\right)}.
\end{equation}
\begin{figure}
\centerline{\includegraphics[width=0.7\linewidth]%
{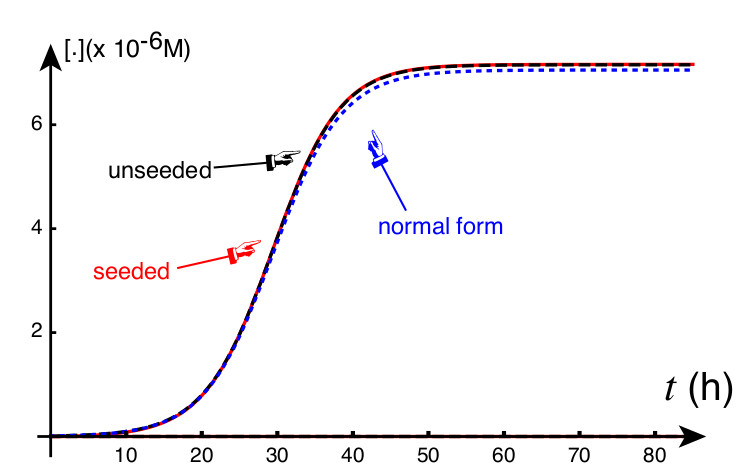}}
\caption{\label{Fig-bif2} %
{Toxic mass concentration $M(t)$ as a function of time for the unseeded (black dashed) system  \eqref{P}-\eqref{m}, for the perfect seeded system (neglecting {\prnuc} and {\sunuc} primary nucleation) (red)  and the normal form approximation of $M(t)$ (dotted blue). The initial conditions were  selected so that  the initial growth rates matched the initial growth rate of the unseeded system. Taking $S = 2.4\times 10^{-13}$ to be a small seed value, the red curve was generated with unseeded  initial conditions; the black dashed 
curve was computed using seeded initial conditions given by 
$(P(0),M(0),m(0)) = (S,S/2,m_0-S)$; and, the blue dashed curve was generated 
by solving \eqref{eqn:nrmfrm:M} with $M(0)=S/2$. Parameters are for  the A$\beta$40 values of Table~\ref{tab:parameters}  and $\lambda=1/2$.}}
\end{figure}
Fig.~\ref{Fig-bif2} shows a comparison of the total toxic mass evolution, 
versus time, obtained for the imperfect unseeded system, the perfect seeded system, and the normal form. As expected, the agreement is excellent as long as the system is close enough to the bifurcation point.

\subsection{Size distribution}
Next, we consider the effect of clearance on size distribution. 
First, we take  $k_{\text {off}}=0$ as suggested by the data sets.
Since, we are interested in the asymptotic size distribution, we can assume that $p_{1}=m_{\infty}$ in Eqs.~(\ref{mbeta1}--\ref{mbeta5}), in which case, we have simply that 
\begin{equation}
p_i= \frac{2 k_{+}m_{\infty} }{\lambda +2 k_{+}m_{\infty} } p_{i-1}=\delta_{0} p_{i-1},\quad \Rightarrow\quad  p_{i}=\delta_{0}^{i-2} p_{2},\qquad i>2.
\end{equation}
Using the definition of $M=\sum_{i>1} i p_{i}$, we obtain:
\begin{equation}
p_{2}=M_{\infty}\frac{  (1-\delta_{0})^{2}}{2-\delta},\quad \Rightarrow\quad  p_{i}=M_{\infty} \frac{  \delta_{0}^{i-2}(1-\delta_{0})^{2}}{2-\delta_{0}} \qquad i>2.
\end{equation}

This analysis is not valid for $\lambda\to0$. In that case, the total mass of the system is systematically transferred to larger and larger particles and in the long-time limit all finite aggregate concentrations tend to vanish and the trivial distribution is $p_{i}=p_{i-1}=p_{2}=0$. However, in that limit, the assumption $k_{\text {off}}=0$ is not justified anymore as even a small value of $k_{\text {off}}$ allows for a non-trivial size distribution. Indeed, with $k_{\text {off}}\not=0$ , we have the following reccurence relation for $c_{i}$
\begin{equation}
0=  - \lambda_i p_i+ 2 k_{+}m_{\infty} (p_{i-1}-p_{i})+2 k_{\text{off}} (p_{i+1}-p_{i}) ,\quad i>2,
\end{equation}
with a single bounded solution fo the form
\begin{equation}
p_{i}={\delta}^{i-2} p_{2},\qquad i>2.
\end{equation}
with
\begin{equation}
\delta=\frac{ k_+
   m_{\infty}}{2 k_{\text{off}}}+\frac{1}{2}+
\frac{\lambda }{4k_{\text{off}}}-\frac{1}{2} \sqrt{\frac{\left(2 k_{\text{off}}+2 k_+
  m_{\infty}+\lambda \right){}^2}{4 k_{\text{off}}^2}-\frac{4 k_+ m_{\infty}}{k_{\text{off}}}}.\end{equation}
An asymptotic expression of $\delta$ for small and large values of $k_{\text{off}}$ gives:
 \begin{equation}\label{approxdelta}
\delta=\begin{cases}&\delta_{0}(1-\frac{2 \lambda  k_{\text{off}}}{\left(2m_{\infty} k_++\lambda
   \right){}^2})+\mathcal{O}\left(k_{\text{off}}^2\right),\quad \text{for}\ \ \quad k_{\text{off}}<m_{\infty} k_{+},
\\
  &\delta_{0}  \frac{2 m_{\infty} k_++\lambda }{2
   k_{\text{off}}}+\mathcal{O}\left(k_{\text{off}}^{-2}\right), \quad\qquad\quad\  \text{for}\ \ \quad k_{\text{off}}>m_{\infty} k_{+}.
   \end{cases}\end{equation}
We see that unless $\lambda=0$, the role of  $k_{\text{off}}$, when sufficiently small, is negligible.
We conclude that clearance (or depolymerization) is sufficient to obtain a non-degenerate size distribution. 
\begin{figure}
\centerline{\includegraphics[width=0.6\linewidth]%
{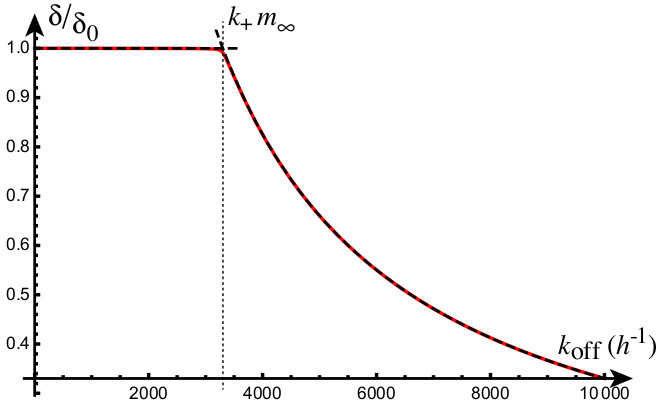}}
\caption{\label{Figdelta} %
{The effect of the parameter $k_{\text{off}}$ on the size distribution can be appreciated by computing $\delta/\delta_{0}$ as a function of  $k_{\text{off}}$. We see that for $k_{\text{off}}<m_{\infty} k_{+} $, the role of $k_{\text{off}}$ is negligible. The dashed curves are given by the asymptotic approximation \eqref{approxdelta}. Parameters are for the A$\beta$40 values of Table~\ref{tab:parameters}  and $\lambda=1/2$.}}\end{figure}

\section{Size-dependent clearance}\label{sec:size-dep-clearance}
Next, we assume that clearance of an aggregate depends on its size. In this case, there is no simple, 
closed equation for the moments, as in Sec.~\ref{sec:size-indep-clearance}, and 
we must study the full system.  
Here, we make a key  assumption about the dependence of the clearance on the aggregate size. 
We assume that there exists  a critical aggregate size, $N$, such that all
{aggregates of size $N$, or greater, are
too large to be cleared. Explicitly, this assumption implies that   $\lambda_{i}=0$, $\forall i\geq N$. We also assume that $k_{\text{off}} = 0$ and 
$n_c=n_2=2$ then \eqref{mbeta1}-\eqref{mbeta5}  can be written
\begin{eqnarray}
&&{\frac{{\text{d}}{{\tilde M}}}{{\text{d}}t}}  
= -\sum_{i=2}^{N-1}\lambda_{i } i p_{i}+2 k_{0}+2 k_{n} p_{1}^{2}+2 k_{+}p_{1} P+2 k_{2}\, \sigma(p_{1}) \tilde M,
 \label{tildeM}\\\label{mp1}
&&  {\frac{{\text{d}}{{p_1}}}{{\text{d}}t}}  
= \lambda_{1}(m_{0}-p_1)-2 k_{0}-2 k_{n} p_{1}^{2}-2 k_{+}p_{1} P-2 k_{2} \sigma(p_{1}) \tilde M,
\\ \label{mp2}
&&  {\frac{{\text{d}}{{p_2}}}{{\text{d}}t}}  
= - {\lambda_{2}\, p_2} +k_{0} + k_{n} p_{1}^{2}-2 k_{+}p_{1} p_{2}+k_{2} \sigma(p_{1}) \tilde M,\\
&&  {\frac{{\text{d}}{{p_i}}}{{\text{d}}t}}  
=  -{\lambda_{i}\, p_i} +2 k_{+}p_{1} (p_{i-1}-p_{i}).\qquad  i>2,\label{mp5}
\end{eqnarray} 
where $P=\sum_{i=2}^{\infty} p_{i}$ and $\tilde M=\sum_{i=2}^{\infty} i p_{i}$.
 The unseeded initial conditions for 
this system are 
\begin{equation}\label{eqn:infinite-unseeded-ic}
p_1(0) = m_0,\quad p_i(0) = 0\,\,\text{for } 2\leq i,\quad \tilde{M}(0) = 0.
\end{equation} 
In general, there is no guarantee of mass conservation.  For instance, if  $\lambda_{i}\leq\lambda_{1}$ $\forall i>2$ and there is at least one $i\geq 2$ 
such that $\lambda_{i}<\lambda_{1}$, then the overall mass of proteins will 
increase in time as shown in Appendix~\ref{masscons}. 


\subsection{{A finite super-particle system}}\label{subsec:size-dep-clearance:superparticle}
{
To study the dynamics of \eqref{tildeM}-\eqref{mp5}, we introduce a finite system 
with equivalent dynamics. Here, we follow \cite{bertsch2016alzheimer} 
(see also \cite{fornari2020spatially}) and introduce 
a super-particle, denoted $q_{N}$, which represents the concentration of all 
aggregates of size greater than or equal to $N$:
\begin{equation}\label{eqn:superparticle-defn}
	q_N = \sum\limits_{i=N}^{\infty} p_i,
\end{equation}
Since $\lambda_i = 0$ for all $i \geq N$; we can take 
the limit of the partial sums of \eqref{mp5} to obtain
\begin{equation}\label{eqn:qn-ode-deriv}
	\frac{{\text{d}}q_N}{\text{d}t} = %
	\sum\limits_{i=N}^{\infty}2 k_+ p_1(p_{i-1} - p_{i}) = %
	\lim\limits_{j\rightarrow \infty} \sum\limits_{i=N}^{j} 2 k_+ p_1(p_{i-1} - p_{i}) = %
	2k_+ p_1 p_{N-1} - 2k_+\lim\limits_{j\rightarrow \infty} p_1p_j.
\end{equation}
Since the monomer concentration $p_1$, remains bounded, for any fixed 
time, the last term of \eqref{eqn:qn-ode-deriv} tends to zero 
as $j\rightarrow \infty$ and the super particle concentration satisfies the equation 
\begin{equation}\label{eqn:qn-ode}
  \frac{{\text{d}}q_N}{\text{d}t} =  2k_+ p_1 p_{N-1}.
\end{equation}
We will distinguish the \emph{finite} system with a super-particle 
from the \emph{infinite} system \eqref{tildeM}-\eqref{mp5} by introducing the notation $q_i = p_i$ 
for $i < N$. Defining $Q=\sum_{i=2}^{N} q_i$, and using \eqref{eqn:qn-ode}, the 
corresponding super-particle system is defined by}
%
%
%
\begin{eqnarray}
&&{\frac{{\text{d}}{{M}}}{{\text{d}}t}}  
=-  \sum_{i=2}^{N-1}\lambda_{i } i q_{i}+2 k_{0}+2 k_{n} m^{2}+2 k_{+}m Q+{2} k_{2}\, \sigma(m) M,
 \label{tM2}\\\label{mq1}
&&  {\frac{{\text{d}}{{m}}}{{\text{d}}t}}  
=\lambda_{1}(m_{0}-m)-2 k_{0}-2 k_{n} m^{2}-2 k_{+}m Q-2 k_{2} \sigma(m) M,
\\ \label{mq2}
&&  {\frac{{\text{d}}{{q_2}}}{{\text{d}}t}}  
= - {\lambda_{2}\, q_2} +k_{0} +k_{n} m^{2}-2 k_{+}m q_{2}+ k_{2} \sigma(m) M,\\
&&  {\frac{{\text{d}}{{q_i}}}{{\text{d}}t}}  
=   -{\lambda_{i}\, q_i}+2 k_{+}m (q_{i-1}-q_{i}),\quad i=2,\ldots,N-1,\label{mqi}\\\label{qN}
&& {\frac{{\text{d}}{{q_N}}}{{\text{d}}t}}  
= 2 k_{+}m q_{N-1}.
\end{eqnarray}The unseeded conditions for \eqref{tM2}-\eqref{qN} are
\begin{equation}\label{eqn:superparticle-unseeded-ic}
	m(0)=m_0,\quad q_i(0) = 0,\,\,\text{for } 2\leq i \leq N,\quad M(0)=0.
\end{equation}
For unseeded initial conditions, the dynamics of the finite system is equivalent to the infinite one in the following sense: First note that $\dot{Q} = \dot{P}$; this follows directly from the 
definition of $Q$, $P$ and \eqref{eqn:superparticle-defn}. Thus, $Q$ and $P$ 
will agree, for all time.  In turn, 
\eqref{tildeM} and \eqref{tM2} coincide when the initial data 
\eqref{eqn:infinite-unseeded-ic} and \eqref{eqn:superparticle-unseeded-ic}, 
respectively, are used; thus $\tilde{M}(t) = M(t)$ in this case.  Finally, 
by definition, $p_i = q_i$ for $2\leq i < N$ and 
\eqref{eqn:superparticle-defn}-\eqref{eqn:qn-ode-deriv} has already established 
that solving \eqref{qN} produces $q_N(t) = \sum_{i=N}^{\infty} p_i(t)$ provided 
the initial conditions agree.  The above establishes an important fact that 
we rely on for the rest of the section; solving \eqref{tildeM}-\eqref{mp5} with 
initial conditions \eqref{eqn:infinite-unseeded-ic} and solving 
\eqref{tM2}-\eqref{qN} with initial conditions 
\eqref{eqn:superparticle-unseeded-ic} yields 
\begin{align}
	m(t)=p_{1}(t), \quad Q(t)=P(t)&, \quad M(t)=\tilde M(t), \label{eqn:infinite-superparticle-system-equivalence}\\ 
	p_{i}(t)=q_{i}(t)\,\,\text{for } 2\leq i < N\quad \text{and}&\quad q_N(t) = \sum\limits_{i=1}^{\infty}p_i(t). \nonumber
\end{align}
We remark, however, that 
$M(t)$, defined as the solution of \eqref{tM2}, is the total toxic mass of both 
\eqref{tildeM}-\eqref{mp5} and \eqref{tM2}-\eqref{qN}, due to 
\eqref{eqn:infinite-superparticle-system-equivalence}, for the unseeded initial 
conditions \eqref{eqn:superparticle-unseeded-ic}; however, $M(t)$ 
cannot be constructed a posteriori from the knowledge of 
$q_i(t)$ where $i=2,3,\dots,N$ in the same manner that $\tilde{M}(t)$ can be 
retrieved from the knowledge of the $p_i(t)$.  That is, we have
$M(t) \neq \sum_{i=2}^{N} i q_i(t).$ Indeed, in the closure process of reducing the full system to a finite one, we lost information regarding the mass of individual 
particles making up the superparticle. Nevertheless,  both the evolution of toxic mass of the full system,  as well as the size distribution (up to size $N$) can be obtained by 
studying the finite system \eqref{tM2}-\eqref{qN}.

\subsection{Toxic mass behaviour}\label{subsec:size-dep-clearance:toxic-mass-behaviour}
Systems such as \eqref{tildeM}-\eqref{mp5} or \eqref{tM2}-\eqref{qN}, with 
size-dependent clearances, do not conserve mass in general 
(see Appendix~\ref{masscons}) and the toxic 
mass may increase with time. We study in more details the particular choice
\begin{equation}\label{eqn:size-dep-clear-sim}
\lambda_i = \lambda / i,\quad\text{for } i=1,2,\ldots,N-1,
\end{equation}
which expresses the modeling assumption that aggregates become increasingly 
difficult to clear as their size increases.  An example of the dynamics of the system 
\eqref{tM2}-\eqref{qN} is shown in Fig.~\ref{FigmM-sizedep}.  
We observe two different behaviors. Initially, up to a time $\tau_{2}$, the system mostly behaves like the conservative no-clearance model ($\lambda=0$) even for large values of clearance. This behavior is markedly different than the one observed in  Fig.~\ref{FigmM-sizeindep2}. Second for larger times, $t> \tau_{2}$, the monomer mass always decreases and the toxic mass always increases as predicted from our general analysis. We observe that larger clearance leads to faster toxic mass creation. This is due to the fact that in healthy homeostasis, production and clearance are balanced. Hence larger clearance implies larger production. The question is then to understand the transition between the two regimes as well as the small and large time behaviors of all species. 
 \begin{figure}[h]
\centerline{\includegraphics[width=0.8\linewidth]%
{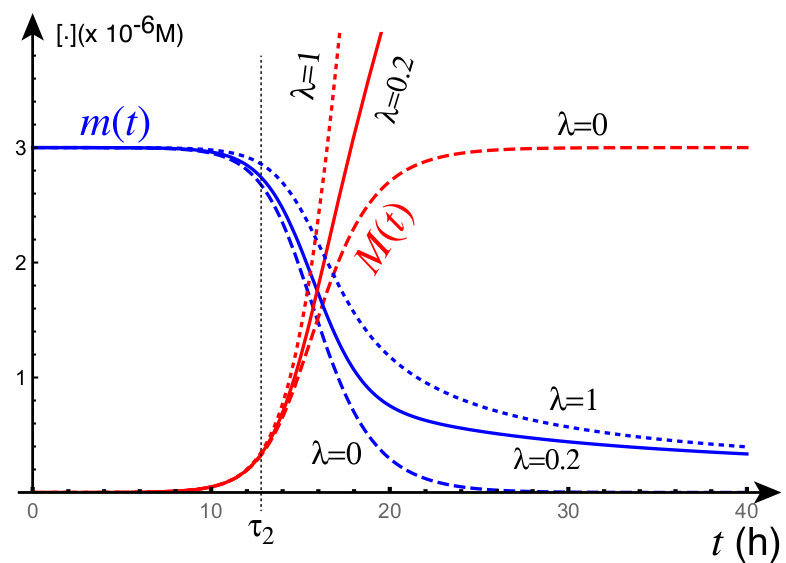}}
\caption{\label{FigmM-sizedep} %
{Toxic mass dynamics for the size-dependent clearance $\lambda_i = \lambda/i$; 
the monomer population concentration ($m(t)$, blue lines) and total toxic mass 
($M(t)$, red lines) are shown for clearance (in h$^{-1}$) 
rates: $\lambda=0$ (dashed), $\lambda=0.2$ (solid), and $\lambda=1$ (dotted). Parameters are for  the A$\beta$40  values of Table~\ref{tab:parameters}  and $\lambda=1/2$ $N=20$}.}
\end{figure}
\begin{figure}[h]
\centerline{\includegraphics[width=0.8\linewidth]{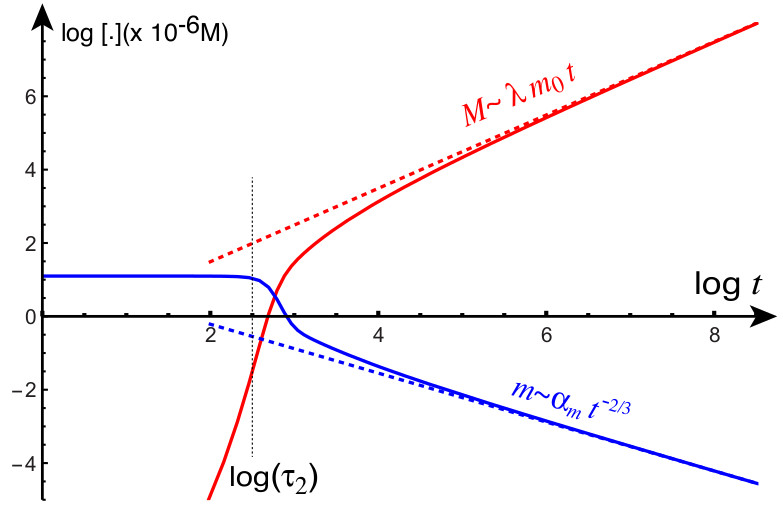}}
\caption{\label{FigmM-sizedep-log} %
Long time (in $h$) concentration (in moles) dynamics of 
\eqref{tM2}-\eqref{qN} ($\lambda_i = \lambda/i$, in $h^-1$) with 
$N=20$ and A$\beta$40 parameters (Table~\ref{tab:parameters}, 
third column); curves for $\lambda = 0.2$ (solid) with asymptotic slopes (dotted). Time runs from 0 to 5000 hours. Parameters are for  the A$\beta$40  values of Table~\ref{tab:parameters} and  $N=20$.}
\end{figure}

\subsection{Long-time dynamics}\label{subsec:size-dep-clearance:long-time-dynamics}
On long time scales, i.e.~long enough so that the monomer concentration begins 
to decrease, the monomer production, aggregation, and nucleation processes 
result in an increase to subsequent toxic species and, therefore, to the overall 
toxic mass $M$.  The asymptotic behavior of the system toxic mass $M$ is observed  
to depend entirely on the production rate, {$\gamma=\lambda m_0$}, as
\begin{equation}
M(t)\mathop{\sim}_{t\to\infty}\gamma\, t.
\end{equation}
This behavior is illustrated in a log-plot in Fig.~\ref{FigmM-sizedep-log}; 
the characteristic time scale, $\tau_{2}$, indicates the time at which the 
monomer mass begins to decay.
Once the asymptotic behavior of $M$ has been established, the equations can be 
balanced asymptotically by the following dynamics:
\begin{equation}\label{eqn:size-dependent:asymptotic-long-time}
m\sim \alpha_{m} t^{-2/3},\qquad q_{N}\sim \alpha_{N} t^{2/3},\qquad q_{i}\sim \alpha_{i} t^{1/3},\quad i=1,\ldots,N-1,
\end{equation}
where the symbol ``$\sim$'' is understood as the long-time asymptotic behavior 
{and the $\alpha_i$ are   constants}. {This asymptotic behavior 
shows} 
that the super-particle dominates the long-term dynamics; {thus} 
$P\sim q_{N}$ for large times. 
Physically, in the long-time limit, the monomer population, renewed by the 
continuous production, is quickly promoted to the super-particle through 
linear aggregation. 
 
\subsection{Early-time dynamics}\label{subsec:size-dep-clearance:early-time-dynamics}
We observe in Fig.~\ref{FigmM-sizedep} that the early-time behavior is not greatly perturbed by altering the clearance rate.
Hence, we can obtain characteristic time scales for {the} amplification of 
the toxic mass by considering the limit $\lambda\to 0^{+}$.
In this case, the early evolution of the toxic mass is  
 governed by the  dynamics 
of \eqref{P}-\eqref{m} with $\lambda = 0$.  %
There are two characteristic time scales of importance.  
First, the time scale  $\tau_1$
associated with {the} exponential growth of the toxic mass in early time  
{via }
the inverse of the positive linear eigenvalue, $\mu=1/\tau_{1}$, {corresponding to} 
the linearization of \eqref{P}-\eqref{m} around the healthy state 
$m=m_{0},\ M=P=0$. The linear eigenvalue is given by 
the positive root of
\begin{equation}
\mu ^2+\mu  \left(4 m_0 k_n-\frac{2 k_2 m_0^2
   K_M}{K_M+m_0^2}\right)-\frac{2 k_+ k_2 m_0^3
   K_M}{K_M+m_0^2}+4 k_+ m_0^2 k_n=0.
\end{equation}
Second, there is a time scale $\tau_{2}$ where  both nucleation and amplification are balanced. It is given by the time for the linearized solution for $M(t)$ to reach $m_{0}$. Hence $\tau_{2}$ is the solution of
\begin{equation}
m_{0}^{2}=\frac{\left(m_0^2 k_n+k_0\right)
   \left(K_M+m_0^2\right)}{2 k_n
   K_M+2 m_0^2 k_n-k_2 m_0 K_M}\left(1-\frac{\text{e}^{{ \tau_{2}}/{\tau_{1}}}}{2}\right).
\end{equation}
For example, for the first parameter set (A$\beta$40) used for the figures, 
these times are $\tau_{1}\approx 1.4$~h and ${\tau_{2}}\approx 12.6$ h. 
{The value of} $\tau_{2}$ is a rudimentary estimate for the time of 
amplification; 
it is a lower bound for the typical time scale of growth 
(see Fig.~\ref{FigmM-sizedep}). Nevertheless, in Fig.~\ref{FigmM-sizedep-log}, 
we see  that $\tau_{2}$ {can indeed act as an} 
indicator for the onset of decay for the monomer mass. A more refined estimate 
can be obtained by using the approximate solution for the full dynamics given 
in \cite{meisl2016molecular}.

\section{The case of a constant free monomer concentration}
\label{sec:const-monomer}
Another interesting case to consider is when the population of monomer is not depleted but remains at a constant level $m_{0}$. We assume that, regardless of other parameters, that \eqref{mbeta1} is 
instead specified by 
\begin{equation}\label{eqn:constant-free-monomer}
\frac{\text{d}p_1}{\text{d}t} = 0,
\end{equation}
so that, with unseeded initial conditions, 
we have $p_1(t) = m_0$ for all time. Assuming again no depolymerization, no fragmentation, and dimer nucleation,   the master equations now read
\begin{eqnarray}\label{mbeta20}
&&  {\frac{{\text{d}}{{p_2}}}{{\text{d}}t}}  
= - \lambda_2 p_2+k_{0} +k_{n} m_{0}^{2}-2 k_{+}m_{0} p_{2}+k_{2} \sigma_{0} M,\label{mbeta21}\\
&&  {\frac{{\text{d}}{{p_i}}}{{\text{d}}t}}  
=  - \lambda_i p_i+2 k_{+}m_{0} (p_{i-1}-p_{i}),\quad i>2,\label{mbeta22}
\end{eqnarray}
where $\sigma_{0}=\sigma(m_{0})$ and $M=\sum_{{i=2}}^{\infty} i p_{i}$ is the total toxic mass.
This is an infinite system of linear ordinary differential equations.  For this system, we 
consider three types of clearance; the size-independent case in addition to two 
different size-dependent paradigms.  All three clearance relations can be 
summarily presented by} a power-law of the form
\begin{equation}\label{eqn:const-monomer-clearance-paradigms}
\lambda_{i}=\lambda i^{\nu}.
\end{equation}
{When $\nu=0$ we recover the size-independent case; when $\nu=-1$ we recover 
the size-dependent diminishing clearance formulation used 
in Sec.~\ref{sec:size-dep-clearance}; and, finally, the case of $\nu=1$ corresponds 
to improved clearance, with increasing size, which could arise due to, for instance, antibody binding.}
Depending on the two parameters $\lambda$ and $\nu$, the solution to this system may have a steady state or increase indefinitely. The question is then to identify the critical values at which this transition happens.

\subsection{{A constant free monomer population with constant clearance}}
{We start with the simple case of constant clearance $\nu=0$; this is the analogue to Sec.~\ref{sec:size-indep-clearance} 
for a constant free monomer assumption (c.f.~\eqref{eqn:constant-free-monomer})} 
{The moments (c.f.~Sec~\ref{sec:size-indep-clearance}) are 
specified by a simple pair of linear equations given by}
\begin{eqnarray}
&&{\frac{{\text{d}}{{P}}}{{\text{d}}t}}  
\,\,= \phantom{\gamma}- \lambda P+ k_{0}+ k_{n} m_{0}^{2}+ k_{2}\, \sigma_{0}  M,
 \label{P2}\\
&&{\frac{{\text{d}}{{M}}}{{\text{d}}t}}  
=\phantom{\gamma}- \lambda M+2 k_{0}+2k_{+}m P+2 k_{n} m_{0}^{2}+{2} k_{2}\, \sigma_{0}  M,
 \label{M2}
\end{eqnarray}
which can be written as
\begin{equation}
\dot{\v q}=A \v q+\v b, 
\end{equation}
where  $\v q=(P,M)^{\text{T}}$,  $\v b=( k_{0}+ k_{n} m_{0}^{2}, 2k_{0}+ 2k_{n} m_{0}^{2})^{\text{T}}$ and 
\begin{eqnarray}
&& A=\left(\begin{array}{cc}-\lambda & a \\b & 2 a-\lambda\end{array}\right)=\left(\begin{array}{cc}-\lambda & k_{2}\sigma_{0} \\2 k_{+}m_{0} & 2 k_{2}\sigma_{0}-\lambda\end{array}\right).
\end{eqnarray}
The constant solution 
{sole steady state} for this system is 
$\v q_{\infty}=-A^{-1} \v b$; $\v q_{\infty}$ is positive and finite if
\begin{equation}\label{eqn:const-monomer:const-clearance:lam-crit}
\lambda>a+\sqrt{a^{2}+a b}=k_{2}\sigma_{0}+\sqrt{k_{2}^{2}\sigma_{0}^{2}+2 k_{2}\sigma_{0} k_{+}m_{0}} = \lambda_{\text{crit}}^{(0)}.
\end{equation}
This condition naturally provides a value for the critical clearance.
Specifically, the largest linear eigenvalue for the system is 
$\kappa=\lambda_{\text{crit}}^{(0)}-\lambda$; 
solutions converge to {$\v q_\infty$} 
exponentially in time (as $\text{e}^{\kappa t}$) for 
$\lambda>\lambda_{\text{crit}}^{(0)}$ and grow unbounded for 
$\lambda\leq \lambda_{\text{crit}}^{(0)}$.  %
The values given in Table~1 for the different parameters show that this estimate is indistinguishable from the case studied in Section~\ref{sec:size-indep-clearance}, which is explained by the fact that at the bifurcation point, the monomer population is constant in both cases.

\subsection{{A constant free monomer population with non-constant clearance}}
We now turn our attention to the general case where the clearance terms are not 
constant. Then, the master equations do not yield a closed system for the moments. 
Nevertheless, due to the simplicity introduced by $p_1(t) = m_0$ being 
constant, we can find conditions for the existence of a fixed-point solution, 
$(p_{2}^*,p_{3}^*,\dots)$ to \eqref{mbeta20}-\eqref{mbeta22}. If such a steady state $p_i^*$ for $i>2$, exists, it must satisfy the recurrence relation
\begin{equation}\label{eq:const-monomer:size-dep:recursive-relation}
p_{i}^*=\delta_{i}p^*_{i-1},\qquad \delta_{i}=\frac{b}{b+\lambda_{i}}=\frac{2 k_{+}m_{0}}{2 k_{+}m_{0}+\lambda_{i}}.
\end{equation}
{%
we note that each of the recursion coefficients, $\delta_i$, is now 
dependent on $i$ via $\lambda_i$.  Define a sequence of real numbers, indexed 
by $i$, as
} 
\begin{equation}\label{eqn:const-monomer:delta-i}
	\Delta_{i} = \prod_{{j=3}}^{i} \delta_{j},\quad i > 2.
\end{equation}
We define $\Delta_2 = 1$ and the $i^{\text{th}}$ steady 
state is  expressible, for all $i\geq 2$, through its recurrence relation as
\begin{equation}
p^*_{i}=\Delta_{i}p^*_{2},\quad i\geq 2.
\end{equation}
Defining
\begin{equation}
\Delta=\sum_{j=3}^{\infty}\Delta_j,
\end{equation}
the  steady state for the total toxic mass solution $M^*$ is then given by
\begin{equation}
M^* = \sum\limits_{i=2}^{\infty} i\Delta_ip_2^* = \Delta p_2^*,
\end{equation}
and an application of \eqref{mbeta21}, at steady state, gives the value of $p_2^*$ as
\begin{equation}
p_{2}^*=\frac{k_{0}+k_{n}m_{0}^{2}}{\lambda_{2}+2k_{+}m_{0}-k_{2}\sigma_{0}\Delta }.
\end{equation}
Therefore, for a fixed point to exist we need the three following conditions to be satisfied
\begin{eqnarray}
&&\text{C1:}\quad \lim_{i\to\infty}  \Delta_{i}=0, \\
&&\text{C2:}\quad \Delta = \sum_{i=2}^{\infty} i \Delta_{i}\ \text{converges},\\
&&\text{C3:}\quad k_{2}\sigma_{0}\Delta -\lambda_{2}-2k_{+}m_{0}>0.
\end{eqnarray}
{%
An analysis of the case $\nu=0$ recovers the previous condition and it can then be verified directly that conditions C1-C3 are satisfied, as expected, for $\lambda> \lambda_{\text{crit}}^{(0)}$.
\subsubsection{Enhanced clearance: $\nu= 1$}
For $\nu=1$, we have (see Appendix C), $\Delta^{(1)}=2+b/\lambda$ and the steady population of dimers, whenever it exists, is given by
\begin{equation}
p_{2}^{*}=\frac{\lambda(k_{0} - k_{n} m_0^2) }{2 (k_{+} m_{0} +\lambda) (\lambda- k_{2}
 \sigma_{0})}.
\end{equation}
Hence, condition C3  leads to  $\lambda>\lambda_{\text{crit}}^{(1)}$ with
\begin{equation}\label{eqn:const-monomer:enhanced-clearance:lam-crit}
\lambda_{\text{crit}}^{(1)}=k_{2}\sigma_{0}.
\end{equation}
We note that the above implies that the critical clearance depends only on the 
secondary nucleation process and, in particular, not the process of elongation 
(c.f.~$\lambda_{\text{crit}}^{(0)}$ in \eqref{eqn:const-monomer:const-clearance:lam-crit}). 
\subsubsection{Reduced clearance: $\nu= -1$}
For $\nu=-1$, the situation is not as simple. The condition C{1} is verified but C{2} leads to  $\lambda>2 b$ for which
\begin{equation}
\Delta^{(-1)}=\frac{\left({\lambda }+2b\right) \left(\Gamma
   \left(\frac{\lambda }{b}-2\right) \Gamma
   \left(\frac{\lambda }{b}+2\right)-\Gamma
   \left(\frac{\lambda }{b}\right)^2\right)}{2 b \Gamma
   \left(\frac{\lambda }{b}\right)^2},
\end{equation}
where $\Gamma(\cdot)
$ is the usual Gamma function. Condition C3 is satisfied if $\lambda>\lambda_{\text{crit}}^{(-1)}$ where $\lambda_{\text{crit}}^{(-1)}$ is the positive solution of
\begin{equation}
f(\frac{\lambda
   }{b})=1+\frac{2 k_{+}m_{0}} {k_{2}\sigma_{0}},\quad\text{with}\quad f(\frac{\lambda
   }{b})=\frac{ \Gamma \left(\frac{\lambda
   }{b}-2\right) \Gamma \left(\frac{\lambda
   }{b}+2\right)}{\Gamma \left(\frac{\lambda }{b}\right)^2}.\label{eaxctlambda}
\end{equation}
This equation always has a solution as $f:z\in [2,\infty]\to f(z)$ is such that $f'(z)<0$, $f(z)\mathop{\to}_{z\to 2} \infty$ and  $f(z)\to_{z\to \infty} 1$. For the parameters listed in Table 1, ${2 k_{+}m_{0}} /{k_{2}\sigma_{0}}\gg1$, in which case, we can approximate the function $f(z)$ close to $z=2$ by $f(z)\approx 6/(z-2)$, which leads to the critical value
\begin{equation}\label{eqn:const-monomer:reduced-clearance:lam-crit}
\lambda_{\text{crit}}^{(-1)}=\frac{8 k_{+} m_{0} (2 k_{2} \sigma_{0}+k_{+} m_{0})}{k_{2}  \sigma_{0}+2
   k_{+} m_{0}}
\end{equation}
This last relation can be further simplified by realizing that $k_{+} m_{0} \gg k_{2} \sigma_{0}$, which leads to 
\begin{equation}
\lambda_{\text{crit}}^{(-1)}=4 k_{+} m_{0}.
\end{equation}
For the parameters given in Table 1, this last approximation of the critical clearance gives the correct value (compared to~(\ref{eaxctlambda})) to 6 digits.
Note that, in contrast to the critical clearance rate for enhanced clearance (c.f.~\eqref{eqn:const-monomer:enhanced-clearance:lam-crit}), 
\eqref{eqn:const-monomer:reduced-clearance:lam-crit} depends only on the elongation rate $k_+$. In particular, in a 
reduced clearance regime, a change in the rate of secondary nucleation has no effect 
on the clearance rate required to keep the system stable.  The general trend that can be 
observed from Table 1 is that $\lambda_{\text{crit}}^{(-1)}>\lambda_{\text{crit}}^{(0)}>\lambda_{\text{crit}}^{(1)}$, as expected.
\subsubsection{Further reduced clearance: $\nu= -2$}
Finally, for $\nu=-2$, skipping computational details, we find that
\begin{equation}
\lim_{n\to\infty} \Delta^{(-2)}_{n}=\frac{1}{4} \pi  \sqrt{\frac{\lambda
   }{b}} \left(\left(\frac{\lambda
   }{b}\right)^2+5 \frac{\lambda
   }{b}+4\right)   \text{csch}\left(\pi  \sqrt{\frac{\lambda
   }{b}}\right),
\end{equation}
which is positive for all finite positive value of $\lambda$. Hence, condition C1 is not satisfied and there is no constant solution or critical value of the clearance that would limit unbounded growth of toxic proteins. We note that we have neglected the effect of fragmentation. For $\nu<0$, the effect of  fragmention is the creation of smaller aggregates that increase the overall expansion of the protein population but also boosts clearance. Indeed since smaller aggregates are more likely to be cleared and we expect a reduction of the critical value of clearance as well as the possibility of a finite value of clearance for $\nu=-2$ or smaller as shown in Meisl \cite{MeislGeorg2016Mpai}.  %
Comparing the different critical clearance values given in Table 1 for the three values of $\nu$, it is clear that that the choice of clearance law has a significant impact on the clearance values as they differ, from the smallest to the largest by 9 orders of  magnitude. Hence, enhancing or inhibiting the clearance mechanism may be extremely important to the overall increase of toxic proteins.

\section{Conclusion}
{%
We have assessed the impacts of production and clearance on the aggregation kinetics 
using a  theoretical model, c.f.~\eqref{mbeta1}-\eqref{mbeta5}, 
that has been experimentally validated \cite{meisl2014differences,cohen2013proliferation,linse2019kinetic}.  Our findings suggest that clearance may mediate toxic aggregation kinetics.  In the case of constant clearance, we showedthat toxic aggregation 
is controlled, directly, by a critical clearance.  Clearance above this level 
provides for a robust environment which is, essentially, free of toxic proteins; 
clearance below this level triggers and instability and a propensity towards 
toxic mass accumulation.  Once toxic aggregation is triggered, the healthy 
monomer population is diminished as aggregates form.  The maximal amount of 
toxic formation is, again, mediated by the clearance level; an effect of the 
mass conservation principle, of this regime.
}

A reasonable in vivo hypothesis is that the clearance may depend on the  aggregates  size $i$. This clearance paradigm has been explored using a simple 
inverse proportionality law $\lambda_i=\lambda/ i$.  The resulting set of equations, for this type of clearance, does not yield a finite system for 
the moments; thus, a super-particle system, with identical trajectories in 
the presence of unseeded initial conditions, has been advanced as a means of 
study.  In the presence of 
any aggregation effects, the system immediately begins accumulating toxic mass; 
even from unseeded initial conditions.  Moreover, mass is not conserved and the 
toxic mass grows unboundedly in time.  The clearance, however, determines the 
asymptotic rate of increase of the toxic mass as a function of time with 
$M(t)\sim \lambda m_0 t$.  The biological implications of a size-dependent 
clearance are quite different than the constant case.  In particular, if clearance 
is size-dependent, results suggest that we have no recourse in halting aggregate 
pathology through enhancing clearance; rather, we can only hope to delay the 
overall trend of toxic accumulation.

The theoretical model of a constant free-monomer concentration was also 
consdired.  This case is particularly interesting 
since, under the assumption of steady states, we see that a notion of critical 
clearance can be established for relations of the form $\lambda_i = \lambda i^\nu$ 
for $\nu\in\left\{-1,0,1\right\}$.  In the case of $\nu=0$, we recover the previous results given for constant clearance. Similarly, in the size-dependent case ($\nu=-1$ and $\nu=1$), there exists a critical value of the clearance so that no aggregation takes place past that value.
Remarkably, our results suggest that, depending on the specific size-dependence, 
the processes of elongation and secondary nucleation contribute to the value of the 
critical clearance to different degrees. An important implication is that, depending 
on the specific mechanism of clearance, inhibition of aggregation should target different 
processes in order to reduce the critical clearance rate.

{%
Overall, the role of  clearance in aggregation 
kinetics is highly non-trivial.  However, our study shows that clearance may play 
an important role in  the aggregation kinetics  of Amyloid-$\beta$ and that additional 
experiments, providing fitted values  for clearance parameters, would serve to 
elucidate appropriate regimes for further study.
}\\

\noindent\textbf{Acknowledgments--\ }
This work was supported by the Engineering and Physical Sciences Research 
Council grant EP/R020205/1 to Alain Goriely and by the John Fell Oxford 
University Press Research Fund grant 000872 (project code BKD00160) to Travis 
Thompson.

\appendix
\section{Normal form for a transcritical bifurcation}\label{appdx:a}
Here we derive the normal form of a transcritical for a general dynamical system. We consider an autonomous $n$-dimensional $\mathcal{C}^2$ vector field of the form
\begin{equation}
\dot{\v x}=\v f(\v x,\lambda),\quad \v x\in \mathbb{R}^n,
\end{equation}
and assume that there exists a constant solution $\v x_0$ such that  $\v f(\v x_0,\lambda)=\v 0$ and a different equilibrium solution in a neighborhood of the critical value $\lambda_0$. The conditions for the existence of a transcritical bifurcation at the critical value $\lambda_0$ are given by Sotomayor's theorem \cite{sotomayor1973generic} and the reduced form the system takes close to that value can be captured by normal form theory \cite{guho83,wi88,go01,go01b}. Here, we use multiple scale analysis to obtain a convenient form of the reduced equations. The result in itself is not original but it may not be obvious to find a direct reference for either the statement or the proof. Therefore, its inclusion may be helpful to the reader.

Using multiple-scale expansion, we expand the solution as
\begin{equation}
\v x=\v x_0+\epsilon\, \v x_1+\epsilon^2\, \v x_2+\ldots,\quad \lambda=\lambda_0+\epsilon \lambda_1.
\end{equation}
where $\v x_0$ is constant and $\v x_i=\v x_i(t,\tau)$, i>1 and $\tau=\epsilon t$ is a slow time \cite{ne74}.
The expansion of the vector field close  to second order is
\begin{align}
f=
&\phantom{+}\v f_0\\
&+\epsilon\left[D\v f_0\cdot \v x_1+\v f_{\lambda,0} \lambda_1\right]\\
&+\epsilon^2\left[D\v f_0\cdot \v x_2+\frac{1}{2} H\v f_0 (\v x_1,\v x_1)+ \lambda_1 D\v f_{\lambda,0}\cdots\v x_1\right]\\&+\ldots,
\end{align}
where $\v f_0=\v f(\v x_0,\lambda_0)$ indicates that $\v f$ is evaluated at the point $(\v x_0,\lambda_0)$ and
\begin{align}
&(D\v f)_{ij}=\frac{\partial f_i}{\partial x_j},&&D\v f_0=D\v f(\v x_0,\lambda_0),\\
&\v f_{\lambda}=\frac{\partial \v f}{\partial\lambda},&& \v f_{\lambda,0}=\v f_{\lambda}(\v x_0,\lambda_0),\\
&(H\v f)_{ijk}=\frac{\partial^2 f_i}{\partial x_j\partial x_k},&& H\v f_0=H\v f(\v x_0,\lambda_0),\\
&(D\v f_{\lambda})_{ij}=\frac{\partial^2 f_i}{\partial x_j\partial\lambda},&& D\v f_{\lambda,0}=D\v f_{\lambda}(\v x_0,\lambda_0).
\end{align}
If the system has a  bifurcation of co-dimension one at $\lambda_0$ then $D\v f_0$ has rank $n-1$ and the following vectors $\v w$ and $\v v$ given by 
\begin{equation}
\v w\cdot D\v f_0=\v 0,\quad D\v f_0\cdot \v v =\v 0,
\end{equation}
define the left and right null spaces of $D\v f_0$.
The generic condition for a transcritical bifurcation to occur is
\begin{equation}
\v w\cdot \v f_{\lambda,0}=0.
\end{equation}
To order $\mathcal{O}(\epsilon)$, the differential equation reads 
\begin{equation}
\dot{\v x}_1=D\v f_0\cdot  \v x_1+ \lambda_1 \v f_{\lambda,0}.
\end{equation}
and we are interested in the solution 
\begin{equation}
\v x_1=c(\tau) \v v,
\end{equation}
whose existence is guaranteed by the condition $\v w\cdot \v f_{\lambda,0}=0$.
To second order $\mathcal{O}(\epsilon^2)$, we have
 \begin{equation}
\dot{\v x}_2+c'(\tau) \v v=D\v f_0\cdot  \v x_2+c^2\frac{1}{2} H\v f_0(\v v,\v v)+ c  \lambda_1 D\v f_{\lambda,0}\cdot\v v.
\end{equation}
The Fredholm alternative gives a condition for the existence of a solution of this inhomogeneous system:
\begin{equation}
\v w\cdot (c'(\tau) \v v)=\v w \cdot\left(c^2\frac{1}{2} H\v f_0(\v v,\v v)+ c  \lambda_1 D\v f_{\lambda,0}\cdot\v v \right),
\end{equation}
which gives the equation
\begin{equation}
c'(\tau)=\beta \lambda_1 c+ \alpha c^2,
\end{equation}
where
\begin{align}
&\alpha=\frac{1}{2}\frac{1}{\v v\cdot \v w}\,\v w \cdot H\v f_0(\v v,\v v)\\
&\beta=\frac{1}{ \v v\cdot \v w}\, \v w \cdot D\v f_{\lambda,0}\cdot\v v.
\end{align}
Taking into account that $\epsilon \lambda_1=\lambda-\lambda_0$ and defining $y=\epsilon c$, the local solution is $\v x=\v x_0+y \v v$ where
\begin{equation}
\dot y=\beta (\lambda-\lambda_0) y+\alpha y^2,
\end{equation}
is the normal form of a transcritical bifurcation at $\lambda=\lambda_0$. The local evolution of the variables for which $v_{i}\not=0$ is given by
\begin{equation}
\dot x_{i}=\beta (\lambda-\lambda_0) (x_{i}-x_{0,i})+\frac{\alpha}{v_{i}}  (x_{i}-x_{0,i})^2.
\end{equation}

\section{Mass balance in the size-dependent clearance case}\label{masscons}
{%
For unseeded initial conditions, we can  show that the total mass of the system is not conserved.
  Assume that, 
for all $2\leq i$ we have $\lambda_i \leq \lambda_1$ and assume that there 
exists some index $j$, with $2\leq j$, such that the inequality is strict 
(i.e.~$\lambda_j < \lambda_1$).  
%
In this case we have
\begin{align}
	\frac{\text{d}\tilde{M}}{\text{d}t} &> -\lambda_1\sum_{i=2}^{N-1}i 
	p_{i}+2 k_{0}+2 k_{n} p_{1}^{2}+2 k_{+}p_{1} 
	P+2 k_{2}\, \sigma(p_{1}) \tilde M \nonumber \\
	& > -\lambda_1\sum_{i=2}^{\infty}i p_{i}+2 k_{0}+2 k_{n} 
	p_{1}^{2}+2 k_{+}p_{1} P+2 k_{2}\, \sigma(p_{1}) \tilde M \nonumber \\
	&= -\lambda_1\tilde{M} + 2 k_{0}+2 k_{n} 
	p_{1}^{2}+2 k_{+}p_{1} P+2 k_{2}\, \sigma(p_{1}) \tilde M. \label{eqn:sizedep-massconsv-loss-a} 
\end{align} 
Likewise for $i=2$ we have a similar inequality  
\begin{align}
{\frac{{\text{d}}{{p_2}}}{{\text{d}}t}}  
&= - {\lambda_{2}\, p_2} +k_{0} + k_{n} p_{1}^{2}-2 k_{+}p_{1} p_{2}+k_{2} \sigma(p_{1}) \tilde M,\nonumber\\
&> - {\lambda_{1}\, p_2} +k_{0} + k_{n} p_{1}^{2}-2 k_{+}p_{1} p_{2}+k_{2} \sigma(p_{1}) \tilde M,\label{eqn:sizedep-massconsv-loss-b}
\end{align}
and likewise for $i>2$.  The above observation shows that the system \eqref{tildeM}-\eqref{mp5} 
grows faster than the constant-clearance case system where $\lambda_i = \lambda_1$ for every $i\in\left\{1,2,\dots\right\}$.
We note that, as in Sec.~\ref{sec:size-indep-clearance},  the total system mass 
for \eqref{tildeM}-\eqref{mp5} is  $\tilde{M}_{\text{tot}} = \tilde{M}+p_1$; this 
follows from the common definition of $\tilde{M}$, here, and $M$ 
(see \eqref{eqn:moments}).  Adding \eqref{tildeM} to \eqref{mp1} and 
using \eqref{eqn:sizedep-massconsv-loss-a} gives 
\begin{equation}\label{eqn:sizedep-massconsv-loss-c}
 \frac{\text{d}\tilde{M}_{\text{tot}}}{\text{d}t} > \lambda_1\left(m_0 - \tilde{M}_{\text{tot}}\right).
\end{equation} 
In the presence of the unseeded initial conditions \eqref{eqn:infinite-unseeded-ic} 
we have that $\tilde{M}_{\text{tot}}(0) = m_0$ so that the left-hand side of 
\eqref{eqn:sizedep-massconsv-loss-c} is strictly positive and mass conservation 
is violated at the outset.  %
Now let $M^{\lambda_1}_{\text{tot}}$ denote the 
total mass of the constant clearance case $\lambda_i = \lambda_1$ for all $i$.  
We know that, in the presence of unseeded initial conditions, a sysetm with constant 
clearance systems conserves mass so that 
\[
\frac{\text{d}M^{\lambda_1}_{\text{tot}}}{\text{d}t} = 0.
\]
From \eqref{eqn:sizedep-massconsv-loss-a} and \eqref{eqn:sizedep-massconsv-loss-b}, which 
holds analagously for $i > 2$ and for $i=1$ we have equality, we can conclude that  
\begin{equation}\label{eqn:sizedep-massconsv-loss-d}
\frac{\text{d}\tilde{M}_{\text{tot}}}{\text{d}t} \geq \frac{\text{d}M^{\lambda_1}_{\text{tot}}}{\text{d}t} = 0,
\end{equation}
for unseeded initial conditions.  Take together, \eqref{eqn:sizedep-massconsv-loss-c} 
implies that the system \eqref{tildeM}-\eqref{mp5}, with unseeded initial conditions, 
initially gains mass while \eqref{eqn:sizedep-massconsv-loss-d} shows that it can 
never lose mass.  Therefore, not only does \eqref{tildeM}-\eqref{mp5} not conserve 
mass but it can never return to the state of initial unseeded mass. 
}

\section{Critical value for enhanced clearance}
For $\nu=1$, the case \eqref{eqn:const-monomer:delta-i} 
takes the form 
\begin{align}
\Delta_i &= \prod\limits_{m=3}^{i}\left(\frac{b}{b+m\lambda}\right) \nonumber\\
&\qquad= \frac{(b+\lambda)(b+2\lambda)}{b^2}\left(\frac{b}{\lambda}\right)^i\left(\left(\frac{b+\lambda}{\lambda}\right)_i\right)^{-1},\quad i\geq3,
\label{eqn:const-monomer:nu-1:delta-i-expression}
\end{align}
where the subscript $(x)_i = x(x+1)(x+2)\cdots(x+i-1)$ denotes ascending factorial 
(i.e.~the Pochhammer symbol).  Defining $\xi = b\lambda^{-1}$ then (C1) is satisfied provided 
\begin{equation}\label{eqn:const-monomer:nu-1:c1-condition}
\lim\limits_{i\rightarrow\infty}\frac{\xi^i}{(\xi+1)_i} = 0.
\end{equation}
}
{%
The function $\xi^i((\xi+1)_i)^{-1}$ is monotonically 
decreasing in both $\xi$ and $i$ and condition C1 is satisfied for any $\xi>0$.%
}
{%
Using $\xi = b\lambda^{-1}$ the expression \eqref{eqn:const-monomer:nu-1:delta-i-expression} 
implies 
\begin{align*}
\Delta^{(1)} &= 2 + \frac{(\lambda+\lambda\xi)(2\lambda+\lambda\xi)}{\lambda^2\xi^2}\sum\limits_{i=3}^{\infty} i \frac{\xi^{i}}{(\xi+1)_i}  \nonumber\\
&= 2 + \frac{(\lambda+\lambda\xi)(2\lambda+\lambda\xi)}{\lambda^2\xi^2}\left(\frac{\xi^3}{2+3\xi+\xi^2}\right) = 2+ \xi.
\end{align*}
Thus we have 
\begin{equation}
\Delta^{(1)} = 2 + \frac{b}{\lambda},
\end{equation}
}
and it follows that $p^*_2$, for $\nu=1$, is determined by the formula
\begin{equation}
p_{2}^{*}=\frac{\lambda(k_{0} - k_{n} m_0^2) }{2 (k_{+} m_{0} +\lambda) (\lambda- k_{2}
 \sigma_{0})}.
\end{equation}

\bibliographystyle{unsrt}

\end{document}